\documentclass[twocolumn,showpacs,preprintnumbers,amsmath,amssymb]{revtex4-1}
\usepackage{epsfig}
\usepackage{graphicx}
\usepackage{epsfig}
\usepackage{color}
\usepackage{rotating}
\usepackage{amssymb}
\usepackage{amsmath}
\usepackage{graphicx}
\usepackage{physics}

\def\beq{\begin{eqnarray}}
\def\eeq{\end{eqnarray}}

\newcommand{\be}{\begin{equation}}
\newcommand{\ee}{\end{equation}}
\newcommand{\bea}{\begin{eqnarray}}
\newcommand{\eea}{\end{eqnarray}}

\begin{document}

\title{A simplified characteristic functions of quantum heat}

\author{Erik Aurell}
\email{eaurell@kth.se}
\affiliation{
KTH -- Royal Institute of Technology,  AlbaNova University Center, SE-106 91~Stockholm, Sweden
}

\begin{abstract}
This paper presents a simplified structure of the characteristic function
of the distribution of energy changes in a bosonic environment interacting
linearly with a quantum system.
This characteristic function of quantum heat can be expressed
using the same functionals (Feynman-Vernon action) as for the quantum state, 
with only the cross-terms in the double path integral shifted in time.
\end{abstract}


\maketitle

\section{Introduction}
\label{sec:Introduction}
The study of fluctuating work and heat in open quantum systems is 
central to the interface
between non-equilibrium statistical physics and quantum information theory, often
referred to as Quantum Thermodynamics~\cite{VinjanampathyAnders2016}.
One approach to such problems is to explicitly model a system of interest interacting
with a bath, and then integrate out the bath, as can be done
explicitly when the bath consists of harmonic oscillators initially in thermal equilibrium.
For the quantum state it was done by Feynman and Vernon in 1963~\cite{FeynmanVernon},
a theory covered in several reviews and monographs \textit{cf.}~\cite{Grabert88,Weiss,BreuerPetruccione}.

One definition of heat in a quantum system is the energy change in a bath.
For the case of harmonic oscillators initially in thermal equilibrium, the expected
value as well as the full generating function can be treated in a manner
closely similar to the Feynman-Vernon approach, as has been investigated by several 
groups~\cite{AurellEichhorn,Aurell2017,Carrega2015,Carrega2016,FunoQuan2017}.

The first goal of this paper is to present a correction to
one expression given in~\cite{Aurell2018b}.
This correction does not change the expected value (quantum thermal power)
which is arguably the most interesting characteristics of a heat flow,
and which has been the focus of many recent studies \textit{e.g.}~\cite{AurellEichhorn,Aurell2017,Aurell2018b,AurellMontana}. 
Though at first glance quite minor, the correction leads 
to a quite considerable simplification in the full generating functions of heat.
The central new result is thus that no new
kernels are needed to describe quantum heat: everything is 
structurally the same as for the quantum state (Feynman-Vernon theory)
with only a shift in the time argument 
in the interaction terms between the forward and backward paths. 
This generalizes two identities to linear order first found in~\cite{AurellEichhorn}.

As the paper is closely related to~\cite{Aurell2018b}
it is organized around re-stating results from that previous paper.
Section~\ref{sec:survey} hence contains a shortened version
of Sections II and Section III of \cite{Aurell2018b}.
The actual calculations are identical to some given in 
Appendix~A in \cite{Aurell2018b}, and are not repeated here.
Section~\ref{sec:newresult}
shows that the corrected result leads to a new and much simpler expression
for the generating function of quantum heat.

\section{The generating function of quantum heat}
\label{sec:survey}
This Section builds extensively on~\cite{Aurell2018b} as well as \cite{AurellEichhorn} and \cite{Aurell2017}. 
The setting is that of one quantum system (``the system'') linearly coupled to a large number of harmonic
oscillators `(``the baths'').
For simplicity I will here consider only one bath. 
The total Hamiltonian of the bath and the system is then
\begin{equation}
H_{TOT} = H_S + H_B + H_I
\label{eq:H_TOT}
\end{equation}
where $H_S$ is the
system Hamiltonian, $H_B$ are the Hamiltonian of the bath, where if appropriate
the Caldeira-Leggett counter-term is included, and $H_I$ is the linear
system-bath interaction.

The generating function of the energy change in the bath is
\begin{eqnarray}
\label{eq:FCS-MAIN}
G_{if}(\nu) &=& \hbox{Tr}_{B}\matrixel{f}{e^{i\nu H_{B}} U e^{-i\nu H_{B}} \left(\dyad{i}\otimes \rho_B(\beta)\right)  U^{\dagger} }{ f}
\end{eqnarray}
where $i$ and $f$ are the initial and final state of the system, 
$\rho_B(\beta)$ is the initial thermal state of the bath at inverse temperature $\beta$
and $\nu$ is the generating function parameter. The probability distribution of
energy changes in the bath, averaged over the initial thermal state, is
the inverse Fourier transform of $G_{if}(\nu)$ \textit{i.e.}
\begin{eqnarray}
\label{eq:FCS-MAIN}
\overline{\hbox{Prob}}_{if}(\Delta E_B) &=& \frac{1}{2\pi}\int e^{-i\nu \Delta E_B} G_{if}(\nu) d\nu
\end{eqnarray}
At zero value of the generating function parameter ($\nu=0$), the generating function reduces
to the Feynman-Vernon transmission probability
\begin{eqnarray}
\label{eq:P-if-abstract}
P_{if}&=& \hbox{Tr}_{B}\matrixel{f}{\left( U \left(\dyad{i}\otimes \rho_B(\beta)\right) U^{\dagger}\right) }{ f} 
\end{eqnarray}
In the Feynman-Vernon approach the two unitary operators and the initial thermal state of the bath 
in \eqref{eq:P-if-abstract}  are 
expressed as path integrals, and then history of the bath is integrated out. 
For baths that are harmonic oscillators this can be done exactly, giving 
\begin{equation}
P_{if} = \int_{if} {\cal D}X {\cal D}Y e^{\frac{i}{\hbar}S_S[X]-\frac{i}{\hbar}S_S[Y]+\frac{i}{\hbar}S_i-\frac{1}{\hbar}S_r} 
\label{P-if}
\end{equation}
where  ${\cal D}X$ and $ {\cal D}Y$ are integrals over the forward and backward paths,  
$S_i$ and $S_r$ are the two terms in the Feynman-Vernon action from 
integrating out the bath, and  $\int_{if}\left(\cdots\right)$ is a short-hand for
projections on initial and final states. 
The Feynman-Vernon action is usually written as products of the sums and differences of the forward and backward paths,
$X+Y$ and $X-Y$. For the following it is convenient to instead write
\begin{widetext}
\begin{eqnarray}
\label{eq:S_i+S_r}
\frac{i}{\hbar}S_i[X,Y]-\frac{1}{\hbar} S_r[X,Y] &=& \frac{i}{\hbar}\int^t \int^s (XX'-YY') k_i(s,s') ds' ds -\frac{1}{\hbar} \int^t \int^s (XX'+YY') k_r(s,s') ds' ds \nonumber \\
                                   && +\frac{i}{\hbar}\int^t \int^s (XY'-X'Y) k_i(s,s') ds' ds +\frac{1}{\hbar} \int^t \int^s (XY'+X'Y) k_r(s,s') ds' ds 
\end{eqnarray}
\end{widetext}
where primed (unprimed) quantities refer to time $s'$ ($s$) and the kernels $k_i$ ($k_r$) are odd (even).
%
%
In \eqref{eq:S_i+S_r} it is only in first double integral where the asymmetric integration limits are necessary.
In the other three terms the integrals in $s$ and $s'$ can be taken symmetric, from initial time up to time $t$. In particular, the cross-terms are
\begin{widetext}
\begin{eqnarray}
\label{eq:S_i+S_r-expr}
\hbox{Cross-terms for $P_{if}$} &=& \frac{i}{\hbar}\int^t \int^t XY' \sum_b C_b C'_b \frac{1}{2m_b\omega_b}\sin\omega_b(s-s') ds' ds \nonumber \\ 
            &&+ \frac{1}{\hbar} \int^t \int^t XY'  \sum_b C_b C'_b \frac{1}{2m_b\omega_b}\cos\omega_b(s-s') \coth(\frac{\omega_b\hbar\beta}{2}) ds' ds 
\end{eqnarray}
\end{widetext}
where the kernels $k_i$ and $k_r$ have been written out
in terms of the two times $s$ and $s'$,  
$C_b$ being the (possibly time-dependent) interaction coefficient between the system and oscillator $b$,
$m_b$ and $\omega_b$ the mass and angular frequency of that oscillator and 
$\beta$ the inverse temperature of the bath.

Let us now consider the generating function $G_{if}(\nu)$ of \eqref{eq:FCS-MAIN}.
Since the path integrals for this quantity are also all Gaussian we must hence have 
\begin{widetext}
\begin{equation}
G_{if}(\nu) = \int_{if} {\cal D}X {\cal D}Y e^{\frac{i}{\hbar}S_S[X]-\frac{i}{\hbar}S_S[Y]+\frac{i}{\hbar}S_i[X,Y]-\frac{1}{\hbar}S_r[X,Y] + {\cal J}^{(2)}(\nu) + {\cal J}^{(3)}(\nu)} 
\label{G-if-path-integral}
\end{equation}
where ${\cal J}^{(2)}(\nu)$ and  ${\cal J}^{(3)}(\nu)$ (in the notation of~\cite{Aurell2018b}) are two new quadratic action terms
symmetric in $s\leftrightarrow s'$:
\begin{eqnarray}
\label{eq:J2-def-MAIN}
{\cal J}^{(2)} &=& \sum_b\frac{i}{2m_b\omega_b\hbar}\int^t\int^t (XY'-X'Y)CC' \sin\omega_b(s-s') \left(\frac{yz'-y'z}{\Delta} -\frac{1}{2}\right) \\
\label{eq:J3-def-MAIN}
{\cal J}^{(3)} &=& \sum_b\frac{i}{2m_b\omega_b\hbar}\int^t\int^t (XY'+X'Y)CC' \cos\omega_b(s-s') \left(\frac{z'-y'}{\Delta} +\frac{i}{2}\coth\frac{\omega_b\hbar\beta}{2}\right)
\end{eqnarray}
\end{widetext}
The auxiliary variables $z$, $z'$,  $y$, $y'$ and $\Delta$ are 
combinations of trigonometric and hyperbolic functions in $\omega_b$, $\nu$ and $\beta$ and given below and
in Appendix~A of~\cite{Aurell2018b}.
The two actions ${\cal J}^{(2)}$ and ${\cal J}^{(3)}$ have been chosen such that they vanish when $\nu=0$, and the
combinations including the two constants ($-\frac{1}{2}$ in \eqref{eq:J2-def-MAIN} and $\frac{i}{2}\coth\frac{\omega_b\hbar\beta}{2}$ in 
\eqref{eq:J3-def-MAIN}) hence cancel with the cross-terms \eqref{eq:S_i+S_r}.
In~\cite{Aurell2018b} the amplitude of ${\cal J}^{(2)}$ was incorrectly given
as $\left(\frac{y'z'-yz}{\Delta} -\frac{1}{2}\right)$.
While this does not change the derivative at $\nu=0$ (which gives expected energy change in the bath),
it changes the full distribution function.
With the correct expression there is a considerable simplification, as we now will see.   

\section{The generating function of quantum heat as a time shift}
\label{sec:newresult}
The starting point is the explicit expressions for the coefficients in \eqref{eq:J2-def-MAIN}
and \eqref{eq:J3-def-MAIN} which were given in~\cite{Aurell2018b} as pertaining to ``Case G'':
$y=\cot(\omega\hbar \nu)$, $y'=\sin^{-1}(\omega\hbar\nu)$,
$z=\cot(\omega\hbar(\nu-i\beta))$ and $z'=\sin^{-1}(\omega\hbar(\nu-i\beta))$,
$\Delta$ being the combination $2(z'y'-yz-1)$. 
From this follows 
$\Delta = 2\sin^{-1}(\omega\hbar(\nu-i\beta))\sin^{-1}(\omega\hbar \nu)\left(1-\cosh(\omega\hbar\beta)\right)$,
$\frac{yz'-y'z}{\Delta} = \frac{1}{2} \cos(\omega\hbar \nu) + \frac{i}{2}\sin(\omega\hbar \nu)\coth(\frac{\omega\hbar\beta}{2})$
and $\frac{z'-y'}{\Delta} = -\frac{i}{2} \cos(\omega\hbar \nu) \coth(\frac{\omega\hbar\beta}{2}) + \frac{1}{2}\sin(\omega\hbar \nu)$.
Let us now rewrite the integrands in
\eqref{eq:J2-def-MAIN} and
\eqref{eq:J3-def-MAIN}. 
For terms proportional to $XY' CC'$ we have
\begin{widetext} 
\begin{eqnarray}
\hbox{Expr.} &=& \sin\omega(s-s') \left(\frac{1}{2} \cos(\omega\hbar \nu) + \frac{i}{2}\sin(\omega\hbar \nu)\coth(\frac{\omega\hbar\beta}{2})\right)
+ \cos\omega(s-s') \left(-\frac{i}{2} \cos(\omega\hbar \nu)\coth(\frac{\omega\hbar\beta}{2}) + \frac{1}{2}\sin(\omega\hbar \nu)\right)\nonumber
\end{eqnarray}
\end{widetext}
By trigonometry this is
$\frac{1}{2}\sin\omega(s-s'+\hbar\nu) 
-\frac{i}{2}\cos\omega(s-s'+\hbar\nu)\coth(\frac{\omega\hbar\beta}{2})$.
The terms proportional to  $X'Y CC'$ are similarly 
$-\frac{1}{2}\sin\omega(s-s'-\hbar\nu) 
-\frac{i}{2}\cos\omega(s-s'-\hbar\nu)\coth(\frac{\omega\hbar\beta}{2})$.
Exchanging labels and including the integrals and the prefactors in
\eqref{eq:J2-def-MAIN} and \eqref{eq:J3-def-MAIN} the cross-terms between the forward and backward paths for the generating function are hence
\begin{widetext}
\begin{eqnarray}
\label{eq:J-2-and-3-expr}
\hbox{Cross-terms for $G_{if}(\nu)$} &=& \frac{i}{\hbar}\int^t \int^t XY' \sum_b C_b C'_b \frac{1}{2m_b\omega_b}\sin\omega_b(s-s'+\hbar\nu) ds' ds \nonumber \\
              &&+\frac{1}{\hbar} \int^t \int^t XY'  \sum_b C_b C'_b \frac{1}{2m_b\omega_b}\cos\omega_b(s-s'+\hbar\nu) \coth(\frac{\omega_b\hbar\beta}{2}) ds' ds 
\end{eqnarray}
\end{widetext}
Comparing to \eqref{eq:S_i+S_r-expr} there is but a simple time shift of the arguments
of the sines and the cosines. 
One consequence of the above is that when the coupling coefficients are time-independent,
any finite derivative with respect to generating function parameter $\nu$ at
$\nu=0$ is equivalent to averaging 
functionals like \eqref{eq:S_i+S_r-expr} with kernels that are time derivatives of the
Feynman-Vernon kernels. To linear order in $\nu$, such an identity was pointed out in~\cite{AurellEichhorn}.

\section*{Acknowledgments}
I thank Prof R. Mulet and the Group of Complex Systems and Statistical Physics, Physics Faculty, University of Havana, Cuba
for an invitation to give lectures at the 2019 School and Workshop where this work was done, and hence
European Union Horizon 2020 research and innovation  programme  MSCA-RISE-2016  under  grant agreement  No. 734439 INFERNET 
for support.

\bibliography{fluctuations}%
\end{document}